\documentstyle[emulateapj]{article}

\lefthead{Yun \& Hibbard}
\righthead{Molecular Gas in Optically Selected Mergers}

\font\smcap=cmcsc10
\newbox\grsign 
\setbox\grsign=\hbox{$>$} 
\newdimen\grdimen
\grdimen=\ht\grsign
\newbox\laxbox 
\setbox\laxbox=\hbox{\raise.5ex\hbox{$<$}\llap
     {\lower.5ex\hbox{$\sim$}}}\ht2=\grdimen\dp2=0pt
\newbox\simlessbox 
\setbox\simlessbox=\hbox{\raise.5ex\hbox{$<$}\llap
     {\lower.5ex\hbox{$\sim$}}}\ht2=\grdimen\dp2=0pt
\def\simless{\mathrel{\copy\simlessbox}}
\def\kms    {\ifmmode{{\rm ~km~s}^{-1}}\else{~km~s$^{-1}$}\fi}
\def\Mo     {~$M_{\odot}$}
\def\hi      {{\smcap H$\,$i}}
\def\Ha      {H$\alpha$}

\def\farcm{\hbox{$.\mkern-4mu^\prime$}}
\def\farcs{\hbox{$\mkern+1.7mu .\!\!^{\prime\prime}$}}

\begin{document}

\title{Molecular Gas in Optically Selected Mergers}

\author{Min~S.~Yun\altaffilmark{1}}
\affil{National Radio Astronomy Observatory,
P.O. Box 0, Socorro, NM~~87801-0387 ({\it myun@nrao.edu})}
\altaffiltext{1}{Present address: University of Massachusetts, Astronomy Department, Amherst, MA~~01003}

\and

\author{J.~E.~Hibbard}
\affil{National Radio Astronomy Observatory, 520 Edgemont Road, 
Charlottesville, VA 22903-2475 ({\it jhibbard@nrao.edu})}

\medskip

\begin{abstract}
We have mapped the 2.6 mm CO $J=1 \rightarrow 0$ emission in three
optically selected ``Toomre Sequence" mergers (NGC~520, NGC~3921,
NGC~4676). The molecular gas distribution is well resolved by the
observations.  For NGC 520 and NGC 4676A, the nuclear gas
concentrations form a disk- or a ring-like structure, and
the gas kinematics are regular and are consistent with 
simple rotation.  Discrete molecular gas complexes are found 
along the stellar bar in NGC~4676B, and the gas kinematics is 
consistent with the disk rotation traced in H$\alpha$. The
molecular gas distribution in NGC~3921 is asymmetric about the stellar
remnant, and both the distribution and kinematics suggest that the
molecular gas has not settled into the center of the remnant.
Molecular gas clouds are detected outside the central regions of
NGC~3921 and NGC~4676, and they may be associated with the tidal tails
and bridges mapped in \hi.  Departures from the canonical scenario for
a merger involving two large spiral galaxies are found in all three
Toomre Sequence mergers studied.  Our data suggest that one of the
progentor disks in NGC~520 and NGC~3921 were relatively gas-poor.
A detailed comparison of these
optically selected mergers and more luminous IR selected mergers is
deferred to a companion paper (Paper II).
\end{abstract}


\section{Introduction}

It is widely believed that a physical collision between a pair of
gas-rich galaxies leads to the concentration of gas and an intense
starburst in the central region of the merger remnant
(\cite{Toomre72,Negroponte83}).  Indeed the majority of infrared
bright galaxies are strongly disturbed systems, and massive
concentrations of molecular gas have been detected in their central
regions (see review by Sanders \& Mirabel 1996 and references
therein).  However, this is not a complete and general picture of all
galaxy collisions since many optically disturbed systems do not show
strongly increased levels of massive star formation (\cite{Bush88}).
The majority of recent studies have focussed mainly on infrared (IR)
selected galaxies, and information on the molecular gas content and
distribution within merging systems of modest infrared luminosity is
sparse in comparison.

%
%
\begin{deluxetable}{lcccc}
\tablewidth{480pt}
\scriptsize
\tablecaption{\sc \small Summary of Observations \label{table1}}
\tablehead{
\colhead{}              &
\colhead{NGC~520N}      & \colhead{NGC~520S}  &
\colhead{NGC~3921}      & \colhead{NGC~4676}  } 
\startdata
Phase Center &&&& \nl
\hskip 0.5in $\alpha$ (1950) &
~01:21:58.8 & ~01:21:59.4 & ~11:48:28.9 & ~12:43:44.5 \nl
\hskip 0.5in $\delta$ (1950) &
+03:32:17.0 & +03:32:09.0 & +55:21:10.0 & +31:00:02.0 \nl
Center Velocity (km s$^{-1}$) & 2260 & 2260 & 5838 & 6600 \nl
Gain Calibrator & 0106+013 & 0106+013 & 1150+497 & 1308+326 \nl
Flux Calibrator & \multicolumn{4}{c}{Uranus, Neptune, 3C 273, 3C 454.3} \nl
Observed Dates &
\multicolumn{2}{c}{9/28/94, 10/3/94} & 10/21/94 & 10/11/94 \nl
 & \multicolumn{2}{c}{12/6/94, 1/26/95} & 2/13/95 & 12/5/94 \nl
 & \multicolumn{2}{c}{2/4/95} & 2/20/95 & 1/12/95 \nl
On-source Integration (hrs) & 11.8 & 11.8 & 19.2 & 16.3 \nl
Sensitivity (mJy beam$^{-1}$) & 12.9\tablenotemark{a} & 
12.9\tablenotemark{a} & 5.8\tablenotemark{b} & 8.1\tablenotemark{b} \nl
$\theta$ (FWHM) (PA) &
\multicolumn{2}{c}{$2\farcs4\times 2\farcs7$ ($-$83$^\circ$)}&
$2\farcs3\times 2\farcs7$ (83$^\circ$) & $2\farcs8\times 3\farcs2$ 
(87$^\circ$) \nl
\tablenotetext{a}{as measured in each 8 MHz (20.8 \kms) channel maps.}
\tablenotetext{b}{as measured in each 16 MHz (41.6 \kms) channel maps.}
\enddata
\end{deluxetable}

A wide range of explanations are possible for the relatively modest
levels of star forming activity observed in the optically selected
colliding galaxies.  Unlike the IR luminous systems, the progenitor
disks of these systems may have been relatively gas-poor.
Alternatively, the progenitors may have been gas-rich, but the initial
conditions of the collisions were such that the bulk of the gas may
have turned into stars or dispersed during the collision (e.g.
\cite{Hos96}).  It is also possible that the IR luminous phase is
generic but brief so that the less luminous mergers are seen in the
pre- or post-burst phase in their evolution.

As a first step in investigating these possible scenarios, we have
used Owens Valley Radio Observatory (OVRO) Millimeter Array to map the
2.6 mm CO $J = 1 \rightarrow$ 0 transition line emission in three
``Toomre Sequence" mergers: NGC~520, NGC~3921, and NGC~4676.  The
``Toomre Sequence" is an optically selected ensemble of strongly
interacting galaxies representing a suggested evolutionary sequence of
disk-disk mergers, based on their stellar tidal tail lengths and the
separation of the two nuclei (\cite{Toomre77}).  Since the members of
this sequence were selected on the basis of optical morphology alone,
they are much less biased towards systems with very high star
formation rates than IR selected samples.

The new high resolution interferometric CO observations allow us to
map the distribution and kinematics of the molecular gas in order to
investigate the response of the molecular material which was
previously distributed in the inner disks of the progenitors.
These data are compared with existing high resolution interferometric
CO observations of IR luminous mergers in a separate paper
(Yun \& Hibbard 2000, hereafter Paper~II). 
In particular, we address whether any
systematic difference exists in the properties of the molecular gas
(which directly fuels the starburst activity) between the less
luminous mergers and the IR selected mergers that may
explain the differences in the IR (and total) luminosity. 

In \S2 we describe the three Toomre Sequence mergers selected for this
study.  The details of the observations and data reduction are discussed
in \S3.  The results are described individually for each system in \S4,
and a discussion of these results follows in \S5.  Finally, in \S6 we
summarize our conclusions.

\section{Sample Selection}

The Toomre Sequence consists of ``Eleven NGC Prospects for Ongoing
Mergers", as sketched and presented by Toomre (1977).  The sequence
was arranged based on the results of simple numerical simulations
conducted by Toomre \& Toomre (1972), which demonstrated the effects
of bound gravitational interactions on the outer regions of disk
galaxies.  This seminal work illustrated that long filamentary
features are the natural consequence of the tidal forces experienced
during such encounters, at the rate of one tail per prograde disk.
Although their numerical technique did not allow for the inclusion of
orbital decay, Toomre \& Toomre (1972) posited that such decay should
occur, leading to the eventual merging of the participants.  This
proposal has been repeatedly confirmed with more sophisticated
numerical treatments (see \cite{BarnesH92}, 1996; \cite{Barnes98} and
references therein).  The natural consequence of such encounters, the
Toomres hypothesized, would be pairs of galaxies spiraling ever closer
together, eventually leaving a single stellar body with two protruding
tidal tails.  The Toomre Sequence was meant to depict this proposed
evolutionary sequence of disk-disk mergers.

Four members of the Toomre Sequence have previously been mapped in CO
with the OVRO interferometer (NGC~4038/9 by \cite{Stanford90,Wilson00},
NGC~520 by \cite{San88b}; NGC~2623 by \cite{Bryant99}; and NGC~7252 by
\cite{Wang92}).  These systems span the entire range of the sequence
from beginning to end.  We chose to improve our understanding of the
dynamical effects occurring during merging encounters by observing
additional systems at the beginning and end of the sequence (NGC 4676
and NGC 3921, respectively) and by re-observing NGC 520.  We selected
these three systems because there are VLA \hi\ mapping and deep broad-
and narrow-band optical observations available to help in the
interpretation of the CO data (from Hibbard \& van Gorkom 1996;
hereafter HvG96).  We chose to re-observe NGC 520 to target explicitly 
the second nucleus, which did not fall within the previously observed
field (\cite{San88b}).  Previous observations and a detailed
description of the three systems are found in HvG96.
Here we summarize their defining optical morphological features and
their location along the Toomre Sequence.  Excellent photographs of
the systems can be found in the Arp ``Atlas of Peculiar Galaxies"
(\cite{Arp66}), and CCD images of the entire systems, included the
extended tidal features, are given in HvG96.  In this paper we will
restrict our figures to the inner few kpc of each merger, as this is
where the CO is concentrated.

NGC~4676 (=``The Mice"=UGC 7938/9=Arp 242=VV 224) is the second member
of the Toomre Sequence, representing an early stage merger where the
two disks are separated by less than one optical diameter in
projection but still distinct.  A bright optical bridge connects the
two disks, and two distinct bright optical tails, each about 50 kpc in
projected length\footnote{All length, mass, and luminosity scales in
the following are derived assuming a value of the Hubble constant of
$H_\circ$= 75 km$\,s^{-1}\,{\rm Mpc}^{-1}$.}, are present.

NGC~520 (=UGC 966=Arp 157=VV231) falls near the middle of the Toomre
sequence (7th of 11), representing an intermediate stage of merging
where two distinct nuclei are seen embedded within a single stellar
body.  The primary nucleus is hidden beneath the prominent dust lane
near the remnant body center, and the secondary nucleus is seen
40$^{\prime\prime}$ (6 kpc) away towards the northwest.  A bright
optical tail stretches 25 kpc to the south, and a broad stellar plume
is seen reaching 60 kpc to the northwest.

NGC~3921 (=UGC 6823=Arp 224=Mrk 430) is the next to last member of the
Toomre Sequence.  It represents the latest stages of a merger, with a
single stellar remnant body exhibiting an $r^{1/4}$ radial light
profile, characteristic of normal ellipticals.  The optical isophotes
are not concentric (Schweizer 1996; hereafter S96), suggesting that
the merger is not fully relaxed.  A stellar tail stretches 65 kpc to
the south, and a broad optical plume reaches 80 kpc to the northeast.

\bigskip

%
%
\begin{figure*}[b!]
\begin{center}
\end{center}
\vspace*{100mm}
\noindent{
\scriptsize \addtolength{\baselineskip}{-3pt} 
{\bf Figure 1.} CO (1--0) spectra of the Toomre
Sequence mergers imaged.  They are obtained by summing all detected fluxes
in the individual channel maps (see Figures~3, 7, \& 11). Single
Dish spectra from the IRAM 30-m telescope ($\theta_{FWHM}$=22$''$)
are shown in dotted lines (NGC~520 [Solomon et al. 1992],
NGC~3921 [Combes, unpublished], NGC~4676 [Casoli et al. 1991]).
The shape and the flux levels agree well between our measurements
and the IRAM 30-m measurements, except in NGC~520 where 
the IRAM 30-m spectrum is significantly lower, possibly due to 
a pointing or calibration problem. \hfil
\addtolength{\baselineskip}{3pt}
\label{fig:spectra}
}\vspace*{-4mm}
\end{figure*}

\section{Observations \label{sec:obs}}

Aperture synthesis CO observations of NGC~520, NGC~3921, and NGC~4676
were carried out with the Owens Valley Millimeter Array between
September 1994 and February 1995.  There are six 10.4 m diameter
telescopes in the array, providing a field of view of about $1^\prime$
(FWHM) at 115 GHz.  The telescopes are equipped with SIS receivers
cooled to 4 K, and typical single sideband system temperatures were
between 300 and 500 K.  Baselines of 15-200 m E-W and 15-220 m N-S
were used, and the details of the observations including the
synthesized beams from naturally weighted data are summarized in
Table~1.  A digital correlator configured with $120 \times 4$ MHz
channels (11.2 \kms) covered a total velocity range of 1340 \kms.
Nearby quasars (see Table 1) were observed at 25 minute intervals to
track the phase and short term instrument gain.  Uranus ($T_B=120$ K),
Neptune ($T_B=115$ K), 3C~273, and 3C~454.3 were observed for the
absolute flux calibration.  The data were calibrated using the
standard Owens Valley array program MMA (\cite{Sco92}) and were mapped
and analyzed using the imaging program DIFMAP (\cite{Shepherd94}) and
the NRAO AIPS software system.  The uncertainty in absolute flux
calibration is about 15\%, mainly due to the uncertainty in
transferring the calibration between the sources and the flux
calibrators.\footnote{Relative accuracy among the measurements
presented here is significantly better, but the absolute uncertainty
is offered because it is more relevant when comparing these
measurements with other measurements.}  The positional accuracy of the
resulting maps is better than $\sim 0\farcs5$.

%
%
\begin{figure*}[tbh]
\begin{center}
\end{center}
\vspace*{105mm}
\noindent{
\scriptsize \addtolength{\baselineskip}{-3pt} 
{\bf Figure 2.} {\it (a)-(c)}
Velocity integrated CO (1--0) map of NGC~4676 at $2\farcs8 \times 
3\farcs2$ ($1.2 \times 1.4$ kpc, PA=87$^\circ$) resolution superposed over the
R-band image (left), H$\alpha$ (middle), and 21 cm \hi\ (right) for
comparison.  An angular separation of 20$''$ corresponds to 8.5 kpc 
at the assumed distance of 88 Mpc.  The FOV of the OVRO dishes at 
115 GHz ($1^\prime$ FWHM) is indicated by a dashed circle. 
The CO contours (white) are drawn at levels 
of 50, 100, 200, 400, 800 $M_\odot$ pc$^{-2}$, and the \hi\ contours 
correspond to column densities of 5, 10, 20, 40, 80 times 
$1\times 10^{19}$ cm$^{-2}$.  \hfil
\label{fig:n4676.p3} 
\addtolength{\baselineskip}{3pt}
}\vspace*{-3mm}
\end{figure*}

A detailed study of all three systems has been conducted in the
optical and 21 cm \hi\ emission by HvG96: radio synthesis observations
of 21 cm \hi\ emission to obtain information on the distribution and
kinematics of extended cold atomic gas; deep broad band $R$ images to
delineate the underlying stellar distribution and any faint optical
tidal extensions; and narrow band {\smcap H$\alpha$+[N$\,$ii]}
observations to reveal regions of current star formation. These data
are compared to the CO distribution and kinematics in the present
work, and are fully described in HvG96.

Additional optical and near infrared (NIR) data on NGC~3921 were
needed in order to examine the dynamical state of this puzzling merger
remnant.  Broadband $BVR$ observations of NGC 3921 were obtained in
January of 1997 with the University of Hawaii 88$^{\prime\prime}$
(UH88$^{\prime\prime}$) telescope. The f/10 re-imaging optics were
used with a Tek2048 CCD, resulting in a plate scale of
0.22$^{\prime\prime}$ pixel$^{-1}$ and a field of view of
7\farcm5. The seeing was $\sim$1\farcs2, and total exposure times of
(1200s, 900s, 600s) were obtained in ($B, V, R$), respectively. The
data were calibrated via observations of Landolt (1983, 1992)
standards observed on the same nights, with resulting zero-point
errors (1$\sigma$) of (0.01 mag, 0.02 mag, 0.03 mag) in $(B,V,R)$.

NIR observations were made at $K^\prime$-band ($\lambda = 2.15 \mu$m,
hereafter referred to simply as $K$, Wainscoat \& Cowie 1992),
obtained with the QUIRC 1024 x 1024 detector on the
UH88$^{\prime\prime}$ telescope in January of 1995.  The f/10
re-imaging optics were used, resulting in a plate scale of
0.187$^{\prime\prime}$ pixel$^{-1}$ and a field of view of
3\farcm2. The observations consist of three 120 sec target-sky pairs,
with the CCD dithered by 1\farcm5 between on-source positions. The NIR
data are uncalibrated.

%
%
\begin{figure*}[tbh]
\begin{center}
\end{center}
\vspace*{170mm}
\noindent{
\scriptsize \addtolength{\baselineskip}{-3pt} 
{\bf Figure 3.} The individual channel maps of CO (1--0) emission in 
NGC~4676 with $2\farcs8 \times 3\farcs2$ ($1.2 \times 1.4$ kpc, 
PA=87$^\circ$) resolution are shown in 41.6 \kms\ velocity increments
upon a greyscale representation of the H$\alpha$ image.
The brightest peak (79.2 mJy beam$^{-1}$) corresponds to
a brightness temperature of 0.60 K above the microwave background.
The contour levels are $-$2, +2, +3, +4, +5,
+6, +7, \& +8 times 9 mJy beam$^{-1}$ ($1\sigma$).  The dominant
CO source is NGC~4676A, and some CO clumps are also seen in 
NGC~4676B.  Several spatially and kinematically coherent
CO features in the region bridging the two galaxies (V = 6516--6641 \kms)
suggest presence of molecular gas in this overlap region. \hfil
\label{fig:n4676ch}
\addtolength{\baselineskip}{3pt}
}\vspace*{-4mm}
\end{figure*}

\section{Results \label{sec:results}}

The observed and derived properties of the sample are summarized in
Table~2 along with other properties of interest.  The far-infrared
luminosities\footnote{$L_{FIR}$ represents the infrared luminosity in
the 40--120 $\mu$m band, calculated from the 60$\mu$m and 100$\mu$m
IRAS fluxes (see \cite{Helou88}).}, $L_{FIR}$, of these three
merging systems range between $(1-5)\times 10^{10} L_\odot$, with a
$L_{FIR}/L_B$ ratio of unity for NGC~4676 and NGC~520 and about 0.2
for NGC~3921.  Therefore these are modest starburst systems at best.
The molecular gas masses\footnote{Standard Galactic conversion of
$N_{H_2}/I_{CO} = 3\times 10^{20}{\rm \,cm^{-2}\,
(K\,km\,s^{-1})}^{-1}$ is used (see \cite{Young91}).}  detected in CO
range between $(1-5)\times 10^9 M_\odot$, accounting for 40\% to 100\%
of the gas masses inferred from single dish measurements.  This is
within a factor of two of the {\it total} molecular gas associated
with our Galaxy ($3.5\times 10^9 M_\odot$; \cite{San84}).  While the
bulk of the molecular gas in our Galaxy is located within an annulus
of 4-6 kpc radius, CO emission in these optically selected merger
systems is concentrated to the central 2 kpc radius, except for
NGC~4676B, whose total molecular gas content and distribution is not
well determined by our data (see below).

The CO (1--0) spectra of each system are produced by summing the
detected emission from each narrow band channel map and are shown in
Figure~1.  All of the CO {(1--0)} spectra show line
widths and shapes comparable to the single dish spectra (shown in
dotted lines; \cite{Casoli91,Sol92}, Combes personal communication).
We do not recover the full line flux measured in the single dish
observations in some cases (see Table~2). Since the interferometer
lacks spacings shorter than about 10 meters, one possibility is that
there is some extended CO emission ($\theta>45^{\prime\prime}$) that
is resolved out in our observations. Another possibility is that some
of the line emission is lost to the limiting surface brightness
sensitivity of the observations.  The peak CO line brightness
temperature observed is only about 0.3-0.6 K for NGC~3921 and NGC~4676
and 5.2 K for NGC~520.  Since the intrinsic CO line brightness should
be at least 10-20 K and may be as high as 30-50 K in starburst
regions, the beam filling factor for the CO emitting regions must be
quite small, less than 10\% and significantly so in some cases.

The molecular gas distribution and kinematics in the individual
systems are discussed in detail below.  For all three observed
mergers, we first compare the CO distribution with the R-band, narrow
band {\smcap H$\alpha$+[N$\,$ii]}, and VLA \hi\ distributions from
HvG96.  We then examine the full CO 3-dimensional kinematic
information as traced by the individual channel maps.  From these
plots it is clear that the CO emitting molecular complexes are well
resolved spatially and kinematically by the aperture synthesis
observations presented here.  We explore the CO kinematics by
comparing the mean velocity and velocity dispersion maps.  The mean
velocity field gives an idea of the large scale orbital motions, and
the intensity-weighted velocity dispersion helps identify the local
centers of the gravitational potential or the sites of large peculiar
velocities.  Finally, we plot the kinematic profiles along the major
axis, and compare these to the 21cm line data and any available
optical kinematics.

%
%
\begin{figure*}[tbh]
\begin{center}
\end{center}
\vspace*{100mm}
\noindent{
\scriptsize \addtolength{\baselineskip}{-3pt} 
{\bf Figure 4.} (a) Velocity integrated CO (1--0) map of NGC~4676 in
linear contours superposed over the H$\alpha$ image.  The contours are 
10, 20, 30, 40, 50, 60, 70, 80, and 90\% of the peak which is 18.4 Jy \kms\
($N_{H_2}= 6.1 \times 10^{22}$ cm$^{-2}$).  
Only weak CO emission is detected along the bar in NGC~4676B. 
The spatial correspondence between the CO and H$\alpha$ is poor,
and extinction by dust associated with the molecular gas (peak
$A_V\sim 120$) offers a natural explanation in the edge-on system
NGC~4676A.  (b) Mean CO emission velocity plotted in contours (in \kms)
superposed over the velocity dispersion (``second moment") map in
greyscale.  The linear greyscale range between 0 (white) to 160 \kms\
(black).  
The velocity gradient is increasing to the north in both galaxies, and thus
the spin orientation of the collision is in a prograde sense. \hfil
\label{fig:n4676mom} 
\addtolength{\baselineskip}{3pt}
}\vspace*{-4mm}
\end{figure*}

%
%
\begin{figure*}[tbh]
\begin{center}
\end{center}
\vspace*{105mm}
\noindent{
\scriptsize \addtolength{\baselineskip}{-3pt} 
{\bf Figure 5.} Position-velocity plot of CO emission in 
NGC~4676A along the kinematic (and morphological) major axis is 
shown in greyscale.  The
molecular gas traces the kinematics of the inner disk with 
a rising rotation curve with a maximum rotation velocity of 270 \kms, 
which flattens at a radius of 960 pc.  The contours 
represent the corresponding P-V plot of the 21cm \hi\ emission (from HvG), 
which is a good indicator of the extent of the outer disk with
a flat rotation curve. They both suggest that the kinematics of 
the inner gas disk ($R<4$ kpc) is relatively undisturbed.  
The S-shaped morphology of the CO emission indicates that the 
molecular gas does not uniformly fill the disk outside 1 kpc radius
and may be confined to a pair of tightly wound spiral arms.  \hfil
\label{fig:n4676lv}
\addtolength{\baselineskip}{3pt}
}\vspace*{-4mm}
\end{figure*}

\subsection{NGC~4676 \label{sec:n4676}}

\subsubsection{Molecular Gas Distribution \label{sec:n4676dist}}

We mapped the CO (1--0) emission in NGC~4676 ($D=88$ Mpc) at $2\farcs8
\times 3\farcs2$ ($1.2 \times 1.4$ kpc) resolution.  The array was
pointed at the region between the two nuclei such that the primary
beam (field-of-view) of the array includes the main bodies of both
galaxies, but very little of the tidal tails (see
Fig.~2a).  Both NGC~4676A (north) and NGC~4676B
(south) are detected in emission, and the integrated CO emission is
contoured in Figure~2.  About 80\% of the CO emission
is associated with NGC~4676A while only weak CO emission is detected
along the bar in NGC~4676B.  The total detected CO flux in NGC~4676A
is $54\pm8$ Jy \kms\ (100\% of the flux measured at the IRAM 30-m
telescope by \cite{Casoli91}), which corresponds to a total molecular
gas mass of $5.5\times 10^9$\Mo\ or about twice as much as in our
Galaxy.  In NGC~4676B we recover only 20\% of the total single dish CO
flux reported by Casoli et al.  Our brightness sensitivity ($\Delta
T_B = 0.08$ K) and a low beam filling factor may explain at least part
of the ``missing" flux.  When the visibility data is tapered to $\sim
5^{\prime\prime}$ resolution, the recovered flux increased to $14\pm2$
Jy \kms, or about 40\% of the total single dish flux.  The CO
morphology does not change substantially in the low resolution maps,
however.  The undetected single dish flux is associated with a
distinct spiky spectral feature occurring between the velocity of
6400-6800 \kms\ in Figure~1.  This feature may 
arise from the molecular gas associated with the
bridging region, poorly represented in our data due to a low beam
filling factor.  Alternatively, this feature may be CO
emission from NGC~4676A picked up by the sidelobe response of IRAM
30-m telescope rather than being intrinsic to NGC~4676B.

The CO emission in NGC~4676A is clearly confined to a nearly edge-on
disk-like structure with a deconvolved size of 1.8 kpc in radius and a
thickness (FWHM) of 250 pc (see Table~2).  
A bridge of stars and gas connecting the two
galaxies is clearly seen in the optical and \hi\ emission (see
Fig.~2c), and several CO clumps are also found in the bridging region
(i.e. at velocities between 6516--6641 \kms\ in
Fig.~3) albeit with relatively low S/N.  Only one such
CO clump appears in the velocity integrated CO map
(Figs.~2a \& ~4a), just southeast of
the main body of NGC~4676A, because only the high signal-to-noise
ratio ($\ge 5\sigma$) features are included in these maps.  The presence
of molecular gas in the bridging region suggests that the disruption
of the inner disks ($R<10$ kpc) has begun in NGC~4676.  Most of the CO
clumps seen in the channel map are unresolved ($\simless1$ kpc in
diameter) with molecular gas masses of $\sim 10^8 M_\odot$.  They are
somewhat larger and more massive than the giant molecular clouds
(GMCs) in our Galaxy, and they may be responding ballistically to the
tidal disruption.  Alternatively, these CO clumps may be molecular gas
condensations forming within the gaseous bridge traced in \hi\ and may
represent possible sites of future star formation.

%
%
\begin{figure*}[t!]
\begin{center}
\end{center}
\vspace*{85mm}
\noindent{
\scriptsize \addtolength{\baselineskip}{-3pt} 
{\bf Figure 6.} {\it (a)-(c)}
Velocity integrated CO (1--0) map of NGC~520 at $2\farcs4 \times
2\farcs7$ ($0.34 \times 0.41$ kpc, PA=$-$83$^\circ$) resolution 
superposed over the R-band image (left), H$\alpha$ (middle), and 
21 cm \hi\ (right) for comparison.  The location and FOV
of the two pointings are indicated by the dashed curves.
The CO emission is not aligned 
with the large scale structure of the optical galaxy, which may be 
the result of accretion of material with angular momentum not aligned 
with the stellar disk.  The CO contours (white) are drawn at levels 
of 50, 100, 200, 400, 800, 1600, 3200 $M_\odot$ pc$^{-2}$. The \hi\ contours 
correspond to column densities of 5, 10, 20, 40, 80, 160, 320 times 
$1\times 10^{19}$ cm$^{-2}$, with dotted contours indicating \hi\
absorption. \hfil
\label{fig:n520.3p}
\addtolength{\baselineskip}{3pt}
}\vspace*{-4mm}
\end{figure*}

One striking feature in the distribution of CO emission in NGC~4676 is
the lack of any correspondence between the brighter \Ha\ peaks
(presumably tracing the present sites of star formation) and the CO
peaks (dense gas concentrations).  Even a marginal anti-correlation is
seen, particularly in Figure~4a which displays a
wider intensity scale for the \Ha\ emission than shown in
Fig.~2b.  The brightest \Ha\ emission in NGC~4676A is
located just outside the southern tip of the CO emitting region, and
significant optical extinction in this nearly edge-on disk offers a
natural explanation -- the peak integrated CO flux of $18\pm3$ Jy
\kms\ beam$^{-1}$ corresponds to $N_{H_2}= 6.1 \times 10^{22}$
cm$^{-2}$ and a mean visual extinction of $A_V\sim 120$ averaged over
the 3$''$ (1.3 kpc) beam.  Even the lowest CO contour in
Fig.~4a corresponds to $A_V\sim 12$.

For the more face-on galaxy NGC~4676B, CO and \Ha\ emission do not
correspond well either.  The brightest \Ha\ peak near the center of
the galaxy has no associated CO emission.  This may reflect the
large extinction associated with the CO emitting clouds. 
Alternatively, this may indicate that
the life time for young stars may be much longer than the cloud
dispersion time scale.  The observed ``twin-peak" CO morphology is
often seen in other barred galaxies, probably due to a bar-driven
dynamical resonance within these disks (\cite{Kenney92}).
Figure~2b shows that both the CO and \Ha\ bars are
offset in the same manner from the underlying optical bar (HvG96).
Similar offsets are seen in hydrodynamical simulations of mergers
(\cite{BarnesH91}) and provide a means of transferring angular
momentum from the gas to the stars, allowing the gas to settle even
deeper into the potential.

Another notable aspect of the CO emission in NGC~4676 is the apparent
contrast in CO luminosity and distribution between the two merging
galaxies.  We detect about 4 times more CO emission in NGC~4676A with
a much more centrally concentrated distribution compared with
NGC~4676B.  The apparent contrast in the CO luminosity in the single
dish measurements by Casoli et al.  (1991) is about a factor of two,
and some of the CO emission associated with NGC~4676B may extend
beyond the inner disk region mapped by us (see above).  One
explanation for the apparent difference in the gas content is that the
progenitor disk for NGC~4676B had less molecular gas.  The two merging
galaxies have similar total \hi\ content (about $3\times 10^9 M_\odot$
each; HvG96), but NGC~4676B appears to be an earlier Hubble type
(SB0/a).  In a survey of molecular gas content among S0 and Sa
galaxies, Thronson et al. (1989) conclude that typical fractional
gas masses in S0's and Sa's are about an order of magnitude less
than those for Sb or Sc spirals, and Young \& Knezek (1898) report
the largest $M(H_2)/M(HI)$ ratios among the S0/Sa Hubble 
types.\footnote{However, the scatter associated with the individual 
Hubble type is substantial in both studies.}
Alternatively, the two progenitor disks started off with similar
amounts of molecular gas but evolved differently under the tidal
disruption because of different internal structure or different
spin-orbit alignment (see \cite{Hos96}).  The difference between these
two scenarios has important consequences for understanding how a gas
disk responds to a tidal disruption.  Our observations alone cannot
distinguish the two however, and this issue needs to be addressed by
future numerical studies.

%
%
\begin{figure*}[b!]
\begin{center}
\end{center}
\vspace*{120mm}
\noindent{
\scriptsize \addtolength{\baselineskip}{-3pt} 
{\bf Figure 7.} The individual channel maps of CO (1--0) emission in 
NGC~520 with $2\farcs4 \times 2\farcs9$ ($0.34 \times 0.41$ kpc, 
PA=$-$83$^\circ$) resolution are shown in 20.8 \kms\ velocity increments
contoured on the greyscale representation of the H$\alpha$ image.
The cross marks the position of the bright radio continuum nucleus.  
The brightest peak (351 mJy beam$^{-1}$) corresponds to
a brightness temperature of 5.2 K.  
The contour levels are $-$3, $-$2, +2, +3, +4, +6,
+10, +15, \& +25 times 15 mJy beam$^{-1}$.  In addition
to the clear rotation signature about the radio nucleus at PA=0$^\circ$,  
there is some evidence for gas associated with the large scale tilted
disk (PA=+45$^\circ$) or perhaps gas associated with the nuclear
blow-out (see channels for V=2153 \& 2195 \kms). \hfil
\label{fig:n520ch}
\addtolength{\baselineskip}{3pt}
}\vspace*{-4mm}
\end{figure*}

\subsubsection{Molecular Gas Kinematics \label{sec:n4676kin}}

NGC~4676A shows a distinct and relatively intact
molecular gas distribution with relatively undisturbed kinematics.
The intensity weighted mean CO velocity is shown in contours
superposed over the greyscale velocity dispersion (``second moment")
map in Fig.~4b.  The velocity dispersion shown in
Fig.~4b exhibit a distinct peak at the center of
NGC~4676A.  This arises from the rapid rise in the CO rotation curve,
and therefore marks the dynamical center of this edge-on disk.  The
major axis position-velocity plot (Fig.~5) shows more
clearly that the CO emitting molecular gas in NGC~4676A is in rotation
about its center, and an apparent flattening of the rotation curve is
hinted by the sudden drop in the velocity gradient outside the central
4$^{\prime\prime}$-5$^{\prime\prime}$ (2 kpc) region.  The
line-of-sight velocity for the CO emission sharply decreases at
$2\farcs5$ (1.1 kpc) radius, forming an S-shaped feature in the
position-velocity plot.  As shown by the solid lines in
Fig.~5, the \hi\ rotation curve remains flat, and this
CO kinematic signature probably does not indicate a real drop in
rotation velocity and may arise from non-circular motions induced by a
non-axisymmetric potential.  A pair of tightly wound molecular spiral
arms can also exhibit a similar kinematic signature.

The peak rotation velocities traced in CO, HI, and \Ha\ are all
about 270 \kms.  The large aspect ratio of the CO emitting region
suggests that the molecular gas disk is viewed within 5-10 degrees of
being edge-on, and the observed peak rotation velocity
should be a good estimate for the true disk rotation speed. 
The derived dynamical mass\footnote{ The dynamical mass estimate
depends on the assumed mass distribution, and a simple rotational
support approximation may result in a slight over-estimation if the
mass distribution is disk-like rather than halo-like (see
\cite{Binney86}).}  $M_{dyn}={{V^2 R}\over{G}}$ inside the 1.8 kpc
radius is then $2.9\times 10^{10}$\Mo, and the molecular gas mass
inferred from the CO emission ($4.9\times10^9$\Mo) accounts for about
20\% of this dynamical mass.  This gas mass fraction is on the upper
end of what is seen in ordinary disk galaxies (a few to 25\%, see
Young \& Scoville 1991 and references therein).

The orbital motion of the molecular gas in NGC~4676B is more difficult
to determine because of its more face-on projection and the patchy CO
distribution.  The observed CO kinematics is consistent with
the \Ha\ rotation curve derived by Mihos et al.~(1993) along the bar
which shows a monotonic velocity gradient consistent with
solid body rotation.  Assuming a disk inclination of 45$^\circ$, the
dynamical mass inside the 4 kpc radius is about $2\times 10^{10}
M_\odot$.  The derived molecular gas mass fraction is then about 7\%,
which is more typical of undisturbed disk galaxies.

The systemic velocities of the two galaxies are very similar
(NGC~4676A slightly more redshifted, by $\sim60$\kms), suggesting that
the two disks are either near their orbital apocenter or moving
mostly in the plane of the sky.  The velocity gradient increases to
the north in both galaxies, consistent with the \hi\ kinematics
reported by HvG96 and the \Ha\ kinematics measured by Stockton (1974)
and Mihos et al.~(1993).  The spin vector of the two disk are aligned
with their orbital motion, and the large tidal tails emerging from
both disks are naturally explained by this prograde spin-orbit resonance
(\cite{Toomre72}).  The CO emitting clouds seen in the bridging region
between the two merging disks have the velocities intermediate between the
systemic velocities of the two disks as expected.

%
%
\begin{figure*}[t]
\begin{center}
\end{center}
\vspace*{80mm}
\noindent{
\scriptsize \addtolength{\baselineskip}{-3pt} 
{\bf Figure 8.} (a) Velocity integrated CO (1--0) map of NGC~520
in contours superposed over the H$\alpha$ image.  The contour levels are 
2, 4, 6, 8, 12, 20, 30, and 50 Jy \kms, and the peak integrated
flux observed is 63.9 Jy \kms ($N_{H_2}= 3.0 \times 10^{23}$ cm$^{-2}$). 
Only the central $5 \times 5$ kpc region surrounding the primary nucleus
of NGC~520 is shown since CO is not detected elsewhere.  
This integrated CO map has a remarkably similar appearance
to that of 1.4 GHz radio continuum map by Condon et al.~(1990), while the
H$\alpha$ emission appears to avoid the CO emission. A larger H$\alpha$ 
image (shown in Figure~6) suggests that
this ionized hydrogen emission is dominated by the bipolar wind blown
out of the intense nuclear starburst region, and the starburst is completely
obscured by the dust associated with the dusty torus mapped in CO (peak
$A_V\sim 600$).  (b) Mean CO emission velocity plotted in contours 
(in \kms) superposed over the velocity dispersion (``second moment") 
map in greyscale.  The linear greyscale range between 0 (white) to 120 
\kms (black).  The well organized velocity field and centrally peaked
velocity dispersion suggests that the molecular 
gas in NGC~520 is in rotation about the primary nucleus. \hfil
\label{fig:n520mom}
\addtolength{\baselineskip}{3pt}
}\vspace*{-4mm}
\end{figure*}

\subsection{NGC~520}

\subsubsection{Molecular Gas Distribution 
\label{sec:n520dist}}

We mapped the CO (1--0) emission in NGC~520 ($D=30$ Mpc) at $2\farcs3
\times 2\farcs7$ ($0.34 \times 0.41$ kpc) resolution, which is a factor
of two improvement over the previous OVRO CO map by Sanders et al. 
(1988; 6$^{\prime\prime} \times 5^{\prime\prime}$).  Two separate
fields were observed, one centered on each of the near-infrared nuclei
(\cite{StanfordB90}), but CO emission is detected only around the main
nucleus (see Fig.~6a).  The $3\sigma$ upper limit for
the molecular gas mass within the 1.3 kpc diameter region surrounding
the second nucleus is $5\times 10^6 M_\odot$ assuming a total line
width of 60 \kms.  This is less than 0.3\% of the molecular gas mass
associated with the main nucleus, and the contrast in the associated
gas mass is quite dramatic.  The absence of an associated molecular
gas complex is somewhat unusual if this were a true stellar nucleus of
a late type galaxy, and we will discuss this point further below (see
\S5).

Nearly all of the $3.0\times 10^9$\Mo\ of molecular gas detected in
the main nucleus position is concentrated in the $1.0
\times 0.4$ kpc (PA=95$^\circ$) disk mapped by Sanders et al.~(1988),
coincident with the bright, extended nuclear radio source
(\cite{Condon90}).  The mean H$_2$ density is about $10^3$ cm$^{-3}$
if the molecular gas is uniformly distributed in a disk with 500 pc
radius and 100 pc thickness, as if the entire molecular complex is a
single super-massive giant molecular cloud.  The full resolution data
results in a detected flux of $285\pm43$ Jy \kms, which is less than
the 325 Jy \kms\ detected by Sanders et al.~(1988).  Smoothing the
data to the 6$^{\prime\prime}$ resolution of the Sanders et al.
recovers a total flux of $404\pm61$ Jy \kms.  The lower resolution
maps do not reveal any additional features in the CO maps. Our
measured flux corresponding to 58\% of the total CO flux measured by
the 14-m FCRAO telescope in its central 45$^{\prime\prime}$ beam area
(\cite{Young95}).  Solomon et al. (1992) report a total integrated
line flux of only 211 Jy \kms\ within the 22$''$ beam of the IRAM 30-m
telescope.  Therefore it is likely that we recovered most of the flux
associated with the nuclear molecular disk, but there may also be
significant systematic errors associated with all of these line flux
measurements compared.

%
%
\begin{figure*}[t]
\begin{center}
\end{center}
\vspace*{100mm}
\noindent{
\scriptsize \addtolength{\baselineskip}{-3pt} 
{\bf Figure 9.} Position-velocity plot of CO emission in 
NGC~520 along the kinematic (and morphological) major axis is 
shown in greyscale. The constant linear gradient suggests that the 
molecular disk in NGC~520 lies
within the solid rotation part of the inner disk. The dark contours 
represent the corresponding P-V plot of 21cm \hi\ emission (dotted
lines represent \hi\ absorption -- from HvG96), and the comparison
with the CO emission suggest that the rotation curve
turns over at the location of the outer radius of the molecular
disk at a 500 pc radius with a maximum rotation speed of 200 \kms.
The brightest CO emission occurs at both extreme velocities,
suggesting that the CO emission decreases towards the center
(see \S\ref{sec:n520kin}). \hfil
\label{fig:n520lv}
\addtolength{\baselineskip}{3pt}
}\vspace*{-4mm}
\end{figure*}

The partial mapping of CO emission by Young et al. suggests that CO
emission extends beyond the nuclear region, along the stellar body of
the galaxy (PA$\sim 45^\circ$). This extended CO emission is not
detected, either because it is resolved out by the interferometer or
because it has insufficient filling factor to be detected by the
brightness limit of the synthesized beam ($\Delta T_B = 0.18$ K).
Evidence for some extended, diffuse gas is seen in the channel maps
(Figure~7, especially channels at 2195 \kms\ and
2153 \kms), and its position angle with respect to the nuclear disk
suggests that the extended gas may be associated with either the large
scale \hi\ disk or with the gas entrained in a galactic superwind
emerging along the minor axis (HvG96).

The anti-correlation between \Ha\ and CO emission is even more
dramatic in the nuclear region of NGC~520.  The clear displacement of
the CO emission from the \Ha\ distribution shown in
Figure~8a suggests that the \Ha\ emission associated
with the nuclear starburst is completely obscured by the dust within
the nuclear gas disk.  The peak integrated CO flux of 63.9 Jy \kms\
beam$^{-1}$ corresponds to a mean column density of $N_{H_2}= 3.0
\times 10^{23}$ cm$^{-2}$ averaged over the $2\farcs5$ (375 pc) beam,
and the inferred large optical extinction ($A_V\sim 600$) is entirely
consistent with the complete obscuration of the nuclear starburst by
the CO emitting clouds.  The observed \Ha\ emission is likely
dominated by the starburst driven ionized wind escaping along the
poles and some scattered light from the nuclear starburst region.  The
1.4 GHz radio continuum map of the nucleus of NGC~520 by Condon et
al.~(1990) is essentially identical in appearance and dimension
(5$^{\prime\prime} \times 2^{\prime\prime}\ {\rm PA=}93^\circ$) and 
is coincident with the CO disk shown in Fig.~8a.  
This is further strong evidence that the vigorous starburst activity
is indeed associated with the molecular gas complex but entirely
obscured. Similarly heavy obscuration of the starburst is found in
other well studied nuclear starburst systems such Arp~220
(\cite{Sco97}).  The observed anti-correlation between CO and
H$\alpha$ not only confirms the diminished H$\alpha$ emission from
massive young stars by extinction (e.g. 
\cite{Bush88,Kennicutt89,Cram98}), but it also casts a serious
doubt on star formation rates inferred from optical or UV tracers for
such starburst systems.

The appearance of the brightest CO emission at the most extreme
velocities seen in the major axis position-velocity plot
(Figure~9) as well as the flatness of the CO emission
along the major axis in the velocity integrated map
(Fig.~8a) suggests that the CO emission does not rise
monotonically inwards but has a central hole, i.e. there is a nuclear
molecular torus rather than a disk (contrast CO contours in
Fig~8a with those for NGC~4676A in
Fig.~4a).  This geometry has some similarity to the
molecular torus found in the nuclear starburst region in M82
(\cite{Lo87,Shen95}), and such a central hole may be a common feature
among nuclear starburst systems (e.g. \cite{Downes98,Carilli98}).

%
%
\begin{figure*}[hbt]
\begin{center}
\end{center}
\vspace*{85mm}
\noindent{
\scriptsize \addtolength{\baselineskip}{-3pt} 
{\bf Figure 10.} {\it (a)-(c)}
Velocity integrated CO (1--0) map of NGC~3921 at $2\farcs3 \times 2\farcs7$ 
($0.87 \times 1.0$ kpc, PA=83$^\circ$) resolution superposed over the 
R-band image (left), H$\alpha$ (middle), and 21 cm \hi\ (right) for
comparison.  An angular separation of 20$''$ corresponds to 7.5
kpc at the assumed distance of 78 Mpc.  
The CO emission appears centered on the remnant on these 
angular scales (15$''$), but it is displaced from the
brightest optical peak as seen in the comparison with H$\alpha$
and as in Figure~12.  
The CO features occur near the ridge of bright
HI emission, but they do not coincide with the \hi\ peaks.  
The CO contours (white) are drawn at levels 
of 50, 100, 200, 400 $M_\odot$ pc$^{-2}$, and the \hi\ contours 
correspond to column densities of 5, 10, 20, 40, 80 times 
$1\times 10^{19}$ cm$^{-2}$. \hfil
\label{fig:n3921.p3}
\addtolength{\baselineskip}{3pt}
}\vspace*{-5mm}
\end{figure*}

%
%
\begin{figure*}[htb]
\begin{center}
\end{center}
\vspace*{150mm}
\noindent{
\scriptsize \addtolength{\baselineskip}{-3pt} 
{\bf Figure 11.} The individual channel maps of CO (1--0) emission in 
NGC~3921 with $2\farcs3 \times 2\farcs7$ ($0.87 \times 1.0$ kpc, 
PA=83$^\circ$) resolution are shown in 41.6 \kms\ velocity increments
contoured on a greyscale representation of the H$\alpha$ image.
The brightest peak (42.1 mJy beam$^{-1}$) corresponds to
a brightness temperature of 0.64 K averaged over the beam area.  
The contour levels are $-$3, $-$2, +2, +3, +4, 
\& +5 times 8 mJy beam$^{-1}$ ($1\sigma$).  The CO emission 
occurs clearly displaced to the northwest of the radio nucleus,
and the gas kinematics is more complex than a simple rotation. \hfil
\label{fig:n3921ch}
\addtolength{\baselineskip}{3pt}
}\vspace*{-4mm}
\end{figure*}

\subsubsection{Molecular Gas Kinematics \label{sec:n520kin}}

The intensity weighted mean CO velocity is shown in contours
superposed over the greyscale velocity dispersion map in
Fig.~8b.  The CO emitting molecular gas in the nuclear
region of NGC~520 shows a smooth, monotonic velocity gradient of 0.36
\kms pc$^{-1}$, which is a clear signature of rotation about its
center.  The regular interval of the iso-velocity contours suggests a
solid body rotation, which is more evident in the major axis
position-velocity plot shown in Fig.~9.  The 21cm
\hi\ absorption (dotted contours) coincides
with that of the CO emission, and this suggests the presence of neutral
atomic gas intermixed with the molecular gas in the nuclear starburst
region.

The CO line widths are represented by the greyscale image in
Fig.~8b.  The broadening of the line width towards the
center of the CO complex suggests that the molecular gas is well
centered on the galaxy nuclear potential. The channel maps and the
iso-velocity contours suggest that the molecular gas disk (or torus)
is well resolved and is close to but not exactly edge-on. Assuming an
intrinsic thickness of 100 pc, we infer an inclination of
$70^\circ-75^\circ$ from the aspect ratio of the CO emitting region.
The peak rotation velocity traced in CO is about 200 \kms\ (210 \kms\
correcting for $i=70^\circ$).  Then the dynamical mass inside 500 pc
radius is $4.9\times10^9$\Mo, and the H$_2$ mass inferred from the CO
emission ($4.3\times10^9$\Mo, see Table 2) nearly entirely accounts
for the total mass inside the 500 pc radius region. This is a larger 
fraction than in NGC~4676A (\S\ref{sec:n4676kin}) and is similarly 
dominant as in the IR luminous mergers such as Mrk~231 and Arp~220
(\cite {Bryant96,Sco97}).  The higher density and temperature
conditions associated with the intense starburst regions may result in
an over-estimation of molecular gas mass if the standard CO-to-H$_2$
conversion is used (\cite{Sco97,Downes98}), and the actual molecular
gas mass traced by CO emission may be smaller by a factor of two or
more.  The orbital period at 500 pc radius is $2\times 10^7$ years,
which is comparable to the time scale for the starburst but two orders
of magnitude smaller than the dynamical time scale for the merger
(HvG96).

One unusual aspect of the nuclear gas torus is that its position
angle and angular momentum vector are misaligned with the larger scale
stellar structures and outer \hi\ disk.  A similar misalignment of
the nuclear disk is seen in other nuclear starburst systems like M82
and may be produced by accretion of inflowing gas with misaligned
angular momentum.  The observed behavior is probably transient in
nature as the continuing infall of mass and resulting torque will
continue to shape the potential.  The systemic velocity of the
molecular gas disk, 2247 \kms, is in good agreement with to that of
the large scale stellar and \hi\ disks (\cite{Stockton80,StanfordB90},
HvG96).

%
%
\begin{figure*}[b]
\begin{center}
\end{center}
\vspace*{75mm}
\noindent{
\scriptsize \addtolength{\baselineskip}{-3pt} 
{\bf Figure 12.} (a) Velocity integrated CO (1--0) map of NGC~3921 in
linear contours superposed over the H$\alpha$ image.  The contours are 10,
20, 30, 40, 50, 60, 70, 80, and 90\% of the peak which is 6.4 Jy \kms\
($N_{H_2}= 3.7 \times 10^{22}$ cm$^{-2}$).  
The bulk of the $1.8 \times
10^9$ \Mo\ molecular gas is concentrated in a single 2 kpc diameter
complex, which is clearly displaced from the optical and radio continuum
peak (marked with a cross).  The HST images by Schweizer et
al.~(1996) reveal an abrupt change in color and an extensive
array of dust filaments in the same region where the CO emission is
detected (peak $A_V\sim 70$ -- see Fig.~13).  
The displacement of molecular gas from
the center of the optical remnant appears real and not an extinction
effect.  (b) Mean CO velocity plotted in contours (in \kms)
superposed over the velocity dispersion (``second moment") map in
greyscale.  The linear greyscale range between 0 (white) to 150 \kms\
(black).  
Unlike NGC~520 or NGC~4676A, there is no clear potential center 
characterized by a large velocity dispersion ($\Delta V > 100$ \kms),
and evidence for reversals in the velocity field is seen.  \hfil
\label{fig:n3921mom}
\addtolength{\baselineskip}{3pt}
}\vspace*{-4mm}
\end{figure*}

%
%
\begin{figure*}[tbh]
\begin{center}
\end{center}
\vspace*{130mm}
\noindent{
\scriptsize \addtolength{\baselineskip}{-3pt} 
{\bf Figure 13.} Comparisons of integrated CO emission in
NGC~3921 with $B$, $K$, and $B-K$ images. A cross in each image marks
the center of NGC 3921 as determined from the HST observations of
Schweizer et al.~(1996). The spatial offset of the central CO complex
from the stellar nucleus is apparent in all continuum band images.
Absence of any K-band emission coincident with the CO feature rules
out the possibility of an obscured nucleus within the molecular gas
complex. A spatial correspondence is found between the CO emission and 
the dust lanes to the northwest (darker $B-K$ colors in the lower
panels). The CO is contoured at 10, 20, ..., 90\% of its peak value. 
North is up and East to the left, with vertical tic-marks drawn every
$1^{\prime\prime}$ and horizontal tic-marks drawn every 0.1 sec. \hfil
\label{fig:n3921K}
\addtolength{\baselineskip}{3pt}
}\vspace*{-4mm}
\end{figure*}

\subsection{NGC~3921 \label{sec:n3921}}

\subsubsection{Molecular Gas Distribution \label{sec:n3921dist}}

NGC~3921 ($D=78$ Mpc) was detected previously in CO (1--0) and CO
(2--1) emission with the IRAM 30 m telescope by Combes and
collaborators (Combes, personal communication).  We mapped the CO
(1--0) emission in NGC~3921 at $2\farcs3 \times 2\farcs7$ ($0.87
\times 1.0$ kpc) resolution, finding the emission confined to at least
two separate complexes (Figure~10): a main complex of 2
kpc diameter near the center of the field, and an unresolved CO
emitting cloud located 7$^{\prime\prime}$ (2.6 kpc) south of the
nucleus. The inferred molecular gas masses are $1.6 \times 10^9$ \Mo\
and $2.0 \times 10^8$ \Mo, respectively.  We recover all of the CO
flux detected at the central position by the 22$''$ beam of the IRAM
30 m telescope (see Fig.~1 \& Table~2), but the CO emission may extend
beyond the area we mapped (Combes, personal communication).  The mean
H$_2$ density for the main molecular gas complex is $\sim$ 150
cm$^{-3}$ if the assumed line of sight thickness is 100 pc.  This mean
density is about 1/6 of that in NGC~520 but comparable to the typical
density in Galactic GMC's.

The integrated CO emission map and the derived mean velocity field are
shown in comparison to the HI and \Ha\ emission in
Figures~10 \& 12.  The locations of
the optical, near-infrared, and radio continuum peak are coincident to
within a fraction of an arcsecond, and the mean position is marked
with a cross.  The main molecular complex in NGC~3921, however, is
displaced by about 2$^{\prime\prime}$ ($\sim$ 760 pc) to the west from
this location.  This suggests that either the gas is physically
displaced from the optical nucleus, or that the true nucleus lies
totally obscured beneath the CO complex (as is the case for NGC~520
and NGC~4676A).

These possibilities are further investigated by comparing the CO
distribution with the optical and NIR (2.15$\mu$m) images, as shown in
Figure~13. The optical and NIR images are found to be
coincident to within a fraction of a pixel ($\sim 0.2^{\prime\prime}$), 
while the peak of the CO emission is displaced by at least 10 pixels
($2^{\prime\prime}$) to the west.  Further, the $K$-band image and
$B-K$ color map shows no hint of a hidden NIR peak beneath the CO
complex. While the peak extinction inferred from our CO observation is
large ($A_V\sim 70$, $A_K\sim 7$), evidence for a second nucleus would
be visible in the NIR image if an extended stellar bulge is
present. It therefore seems that the observed displacement between the
molecular gas and the stellar nucleus is real. The CO emission appears
to be associated with a concentration of dust, as indicated by the
dark greyscales in the $B-K$ map in Fig.~13.

A comparison between the gas distribution and the HST Planetary Camera
(PC) image of NGC 3921 by Schweizer et al.~(1996; see especially their
Figure~4) supports this picture. This comparison is shown in
Fig.~14, where the CO emission is contoured upon the
$V$-band PC image of NGC 3921 of Schweizer et al. and upon the same
image after a best-fit model light distribution has been subtracted.
In the latter image dust lanes show up in white, and we confirm the
conclusion reached above that the main CO complex coincides with an
intricate system of dust lanes concentrated mainly to the west of the
nucleus. Such a displacement is unexpected since the highly
dissipative nature of the cold gas should cause it to
settle into the gravitational potential within an inner dynamical time
of $\sim 10^7$ years.  The non-axisymmetric appearance of the dust
lanes gives the impression that the gas continues to spiral in towards
the center, and the gas may be still settling within the merger
potential.  The combination of off-centered or ``sloshing" optical
isophotes (\cite{Schwe96a}) and non-axisymmetric dust lanes and
disordered kinematics led Schweizer et al.~(1996) to suggest that ``on
scales of order 100--1000 pc the gas and dust in NGC~3921 do not form
a well settled nuclear disk".  The displaced CO appears to be a strong
confirmation of this interpretation.

The molecular gas clump located to the south of the main complex lies
along the ridge of optical, \Ha, and \hi\ emission that delineates the
beginning of the southern tail (see Fig.~10, and also
HvG96).  This raises an intriguing possibility that this feature is
associated with the gas-rich southern tail.  Tidal features are
generally very difficult to detect in CO (e.g. \cite{Smith94,Smith99}),
but molecular gas has been found in the bridging regions in NGC~4676
(see \S\ref{sec:n4676dist}) and in Arp~105 ({\cite{Braine00}).
The continued infall of gas from the outer regions might offer an
explanation for the displaced molecular complex in NGC~3921 (see \S5).

%
%
\begin{figure*}[tbh]
\begin{center}
\end{center}
\vspace*{80mm}
\noindent{
\scriptsize \addtolength{\baselineskip}{-3pt} 
{\bf Figure 14.} Comparisons of integrated CO emission in
NGC~3921 with (left) the $V$-band HST image from Schweizer et al.
(1996, their Figure 4a), and (right) with the same image, after a
best-fit model light distribution has been subtracted (their Figure
4b). The region shown is approximately the same as in
Fig.~13, although the field-of-view has been rotated.
The model-subtracted image shows much more clearly the dust lanes
concentrated to the west of the system, which appear white in this
figure. The CO is contoured as in Fig.~13, and is seen
to be very closely associated with the dust lanes to the west. The
chaotic appearance of the dust lanes led Schweizer et al.~(1996) to
suggest that NGC 3921 is still in the process of forming a nuclear
disk. The displaced CO appears to be a strong confirmation of this
interpretation. \hfil \label{fig:n3921HST}
\addtolength{\baselineskip}{3pt}
}\vspace*{-4mm}
\end{figure*}

\subsubsection{Molecular Gas Kinematics \label{sec:n3921kin}}

The CO channel maps for NGC 3921 are given in Fig.~11. 
These maps show a general southwest-to-northeast velocity gradient
along the main CO complex.  However, numerous peaks and extensions
appear and disappear in different channels at various position
angles.  The southwest complex appears at velocities between 5752--5838
\kms, but with a velocity gradient opposite in sense to the main
complex (i.e., decreasing velocities from south to north).  The
intensity weighted mean CO velocity, shown as contours superposed over
a greyscale of the velocity dispersion map in Fig.~12b,
also show the same gradient along the main CO complex and the velocity
reversal in the SW complex.  On the other hand, the local maximum in
velocity dispersion characteristic of a local gravitational potential
well is not seen in Fig.~12b, supporting the
conclusion reached earlier that no distinct potential maximum exists
within the main molecular gas complex.

Towards the main body, the \hi\ column density drops rapidly, and the
\hi\ tail cannot be traced as an individual kinematic structure.  The
\hi\ which appears projected onto the main body is spread over a broad
range of velocities, similar in width to the OVRO and single dish
profiles ($\sim$ 460 \kms, Table 2) and the stellar velocity
dispersion ($\sim$ 280 \kms; \cite{Lake86}).  Unlike the case for the
other two systems, the molecular gas mass accounts for a relatively
small fraction of the dynamical mass in the inner regions (about 10\%
and 5\% inside a 1 and 2 kpc radius, respectively).

The position-velocity plot of the CO data is not very informative, so
we instead illustrate the CO kinematics by comparing them to the
optical emission line kinematics of S96, which were taken through a
2$^{\prime\prime}$ slit centered on the optical nucleus at a PA of
45$^\circ$. This comparison is presented in Figure~15.
The emission line kinematics are in general agreement with that the
CO, suggesting that the two are tracing the same feature.  To the SW,
both the optical emission lines and CO show a sudden velocity reversal
to lower velocities, showing that neither component clearly traces a
simple rotation about the central potential.  S96 derives a systemic
velocity of 5926 \kms\ from the emission lines, while the CO data may
suggest a somewhat lower value. However, given the lack of a
recognizable systematic rotational signature, we believe the data are
inconclusive as to this point.

The observed velocity reversals are very similar to those seen in the
well studied merger remnant NGC~7252. A comparison between the \Ha\
(\cite{Schwe82}), CO (\cite{Wang92}), and tidal \hi\ kinematics in
that system (\cite{Hib94,Hib95b}) show that the velocity reversals
occur where tidal features are projected near the inner regions, and
can be explained by supposing that strong streaming motions of
infalling gas from the tidal regions are superposed upon the
kinematics of a central molecular disk.  In NGC 3921 the velocities of
the southern clump (encircled ``S'' in Fig.~15) appear
at similar velocities as the blueshifted ionized gas. These velocities
are also similar to the \hi\ velocities at the base of the southern
tail (HvG96), and this kinematic component may be associated with
infalling gas from the tidal regions.

In summary, the inner region of NGC 3921 has an off-center
molecular gas complex which has not settled into a disk,
and a second component of gas may be infalling from the tidal regions.
This is rather different from the nuclear molecular complexes
seen in most single nucleus mergers, a point we will return to below.

\section{Discussion}

The majority of mergers mapped in CO show a central molecular disk
with well defined rotational kinematics
(\cite{Sco97,Downes98,Bryant99}).  Early gas inflow and the formation
of a compact nuclear gas complex, even before the coalescence of the
two stellar nuclei, are also seen in numerical studies of galaxy
merger (e.g. \cite{BarnesH96}).  This is a direct consequence of the
fact that gas can shock and radiate energy away.

While some of the observed properties of these three Toomre Sequence
mergers follow this canonical picture of a merger involving two
large spiral galaxies, significant departures are also found.
A dense concentration of molecular
gas is associated with the central nucleus of the late stage merger
NGC~520, but no other molecular gas concentration is detected on
the second nucleus or along the region bridging the two nuclei.
The formation of a compact nuclear gas complex appears to be
well underway in the merging disk of NGC~4676A, but evidence for gas
inflow and formation of a central gas concentration
is not evident in NGC~4676B.
In NGC 3921, the molecular gas complex is significantly displaced
from the peak of the optical, near-infrared, and radio continuum
emission ($\sim$ 760 pc).

One possible explanation for these departures from the canonical
merger scenario is that one or both progenitor disks of these
optically selected merger systems may have been relatively gas-poor.
Based on the analysis of the stellar and \hi\ tidal features, Hibbard
\& van Gorkom (1996) suggest that both NGC~520 and NGC~3921 are the
products of an encounter between one gas-rich disk and one gas-poor
system, such as an S0 or Sa galaxy (e.g. Thronson et al. 1989). 

This suggestion for NGC~3921 is further supported by the observations
of Schweizer et al.~(1996), who find many fewer young globular
clusters in this system compared to the gas rich mergers NGC 4038/9
and NGC 7252.  The absence of molecular gas directly associated with
the second nucleus of NGC~520 and the stellar nucleus of NGC~3921 is
then naturally explained if little molecular gas has been funneled
into the nuclei during the merger process.  The absence of gas in one
progenitor also lessens the frequency of gas cloud collisions in the
second disk, especially within the central few kpc (\cite{Olson90}),
and therefore less dissipation and angular momentum transport is
expected.

The presence of two large, atomic gas-rich tidal tails associated with
NGC~4676 suggests that both progenitor disks were gas-rich (see
HvG96), and a different explanation may be needed.  As stated
earlier, NGC~4676B appears to be an early Hubble type spiral and
may also have been relatively molecular gas poor initially.
A numerical study by Mihos \& Hernquist (1996) has shown that the
presence or absence of a massive bulge in the progenitor disk can
produce a large difference in the evolution of the gas distribution
and kinematics, the peak gas density achieved, and the resulting star
formation activity.  The apparent difference between the
current molecular gas content in the two merging disks in NGC~4676 may
be explained more naturally by differences in the internal
structure of the progenitors. 

While it is significant that these three optically selected mergers
show notable departures from the canonical merger scenario, we should
also keep in mind that they nevertheless represent only three
individual snapshots of their respective merger evolutions.  For
example, the apparent difference in the status of molecular gas
between the two merging disks in NGC~4676 may simply reflect a slight
delay in the onset of the gas inflow created by the difference in the
spin-orbit coupling, however fortuitous this may seem.  Similarly,
NGC~3921 may be in a brief, early stage in its merger evolution and
may soon develop a molecular disk.  Schweizer (1996) has also
presented observation evidence that NGC 3921 is a dynamically young
merger. However, central CO disks are commonly found in both earlier
(e.g. NGC 520 and NGC 4676A) and later stage (e.g. NGC 7252) mergers.
One may further postulate that NGC~3921 represents a stage even
further in its evolution where the original central gas complex is
already dispersed.  In this scenario, the observed gas in the central
region is largely the result of recent inflow from the tidal tails
(see Hibbard \& Mihos 1995).

Finally, a clear and significant quantitative result from this imaging
study of optically selected merger systems is that their central
molecular gas surface density is systematically smaller by orders of
magnitude compared with the IR selected mergers (see Table~2).  
Further, the level
of activity associated with each of the merger systems appears to
depend strongly on the central gas density.  We will discuss this and
other comparisons of global properties between the optically selected
and IR selected mergers in a separate paper (Paper II).

%
%
\begin{figure*}[tbh]
\begin{center}
\end{center}
\vspace*{75mm}
\noindent{
\scriptsize \addtolength{\baselineskip}{-3pt} 
{\bf Figure 15.} Position-velocity plot of CO emission in
NGC~3921 in comparison to the kinematics of the optical emission 
lines is shown on the right panel while the actual positions of 
each sampled component are shown on the right panel.  
The \Ha\ kinematics of S96 are indicated by filled
circles. CO velocities measured from the first moment map across the
peak of the CO distribution at a PA of 45$^\circ$ are shown as small
``+'''s.  The ``clean components'' associated with the southern clump
are denoted by encircled ``S".  The size of the symbol is 
proportional to the S/N of the component. \hfil
\label{fig:n3921lv}
\addtolength{\baselineskip}{3pt}
}\vspace*{-4mm}
\end{figure*}

\bigskip

\section{Summary}

In an effort to better understand the nature of the luminous infrared
galaxy phenomenon and to scrutinize the details of the merger
scenario, we have mapped the CO emission in an optically selected
sample of on-going mergers with moderate IR luminosity (``Toomre
sequence", \cite{Toomre77}) at a resolution and sensitivity comparable
to those of the more IR luminous objects, and the response of the
molecular gas material deep inside the potential of the colliding
galaxies is traced.  The summary of our findings is as follows:

1. A compact ($R\le2$ kpc) molecular complex is found well centered on
the inner stellar disk of NGC~4676A, forming a disk- or a ring-like
structure with the regular kinematics consistent with simple rotation.
This molecular gas concentration contributes a significant fraction
(20\%) of the total mass in the nuclear region.  The CO emission
occurs along the 7 kpc stellar bar in NGC~4676B.  The surface
brightness of the CO emission is extremely low, and tapering the data
from 3$''$ to 5$''$ resolution nearly doubles the detected flux.

2. Nearly all of the CO emission detected at the primary nucleus
position in NGC~520 is concentrated in a 1 kpc diameter ring-like
structure, coincident with the extended radio continuum source mapped
by Condon et al. (1990).  The derived H$_2$ mass and the dynamical
mass are comparable, and the gas mass must constitute a large fraction
of the total mass in the nuclear starburst region.  No CO emission is
detected on the second nucleus in NGC~520.

3. The molecular gas distribution in NGC~3921 is quite different from
any other singly nucleated merger mapped in CO.  The CO emission is
significantly displaced from the peak of the optical, near-infrared,
and radio continuum emission ($\sim 760$ pc), and both the
distribution and kinematics suggest that the molecular gas has not
settled into the new remnant potential. The molecular gas complex
contributes a minor fraction ($\simless10$\%) of the dynamical mass 
in the inner regions.

4. Molecular gas clouds are detected outside the central regions of
NGC~3921 and NGC~4676.  They may be associated with the tidal tails
and bridges mapped in \hi.

5. A consistent trend of anti-correlation is seen between CO and
H$\alpha$ emission in NGC~520 and NGC~4676A, and the large extinction
inferred from the CO emission ($A_V=$ 600 \& 120) suggests that the
intense nuclear starburst regions in these galaxies are completely
obscured.  This finding offers a natural explanation for the severe
under-estimation of the star formation rate by optical tracers
compared with the IR (e.g. \cite{Bush88}, Kennicutt 1989,1998).

6. Departures from the canonical scenario for a merger involving two
large spiral galaxies are found in all three Toomre Sequence mergers
studied.  A relatively gas-poor progenitor, as inferred from the
optical and \hi\ observations by HvG96, offers a possible explanation
for NGC~520 and NGC~3921.  A gas-poor progenitor disk or rapid
evolution in gas content by some dynamical process (i.e. \cite{Hos96})
may offer a plausible explanation for the low molecular gas density in
NGC~4676B.}

7. The central molecular gas density in these optically selected
mergers is systematically smaller by an order of magnitude compared
with the IR selected mergers, the level of activity associated with
each of the merger systems appears to depend strongly on the central
gas density.  This and other detailed comparisons of global properties
between the optically and IR selected mergers will be presented in
Paper~II.

\acknowledgements

The authors thank F. Combes for discussing the results of her IRAM
30-m CO (1--0) observations of NGC 3921 prior to publication and
allowing us to show their spectrum in Figure~1, and
F. Schweizer for sending us his \Ha\ kinematic data on NGC 3921 and
for a critical reading of an earlier version of this manuscript.  We
thank J. Barnes, M. Rupen, N. Scoville, J. van Gorkom, and
J. Wang for useful discussions.  We also thank the anonymous referee,
whose helpful suggestions improved this manuscript.  This research was
supported in part by NSF Grant AST 93-14079.  
The National Radio Astronomy Observatory is a facility 
of the National Science Foundation operated under cooperative agreement
by Associated Universities, Inc.

\newpage

%
%
\begin{deluxetable}{lcccc}
\tablewidth{480pt}
\scriptsize
\tablecaption{Summary of Observed and Derived Properties \label{table2}}
\tablehead{
\colhead{}               & \colhead{NGC~4676A}      &
\colhead{NGC~4676B}      & \colhead{NGC~520}  &
\colhead{NGC~3921}       } 
\startdata
RA (1950) & ~12:43:44.1 & ~12:43:45.3 & ~01:21:59.6 & ~11:48:28.9 \nl
DEC (1950) & +31:00:19 & +30:59:47 & +03:31:53 & +55:21:23 \nl
Distance (Mpc)\tablenotemark{a} & 88 & 88 & 30 & 78 \nl
$<V_{CO,hel}>$ (km s$^{-1}$) & 6632$\pm$42 & 6590$\pm$42 & 2247$\pm$21 & 
5880$\pm$42 \nl
$\Delta V_{CO,FWZI}$ (km s$^{-1}$) & 546$\pm$42 & 252$\pm$42 & 
499$\pm$21 & 462$\pm$42 \nl
Deconvol. Size (PA)
 & $8\farcs5\times 1\farcs1$ ($5^\circ$) & $16^{\prime\prime} \times 
   2^{\prime\prime} (22^\circ)$ & $7\farcs1\times 2\farcs9\ (95^\circ)$ 
 & $5\farcs0\times 2\farcs3$ ($38^\circ$) \nl
 & ($3.6\times 0.5$ kpc) & ($7\times <1$ kpc) &
   ($1.0\times 0.4$ kpc) & ($1.9\times 0.9$ kpc) \nl
$\Delta T_{CO,peak}$  &
$0.60\pm0.09$ K  & $0.28\pm0.04$ K  & $5.2\pm0.7$ K  & $0.64\pm0.09$ K  \nl
$S_{CO}\Delta V$ (Jy km s$^{-1}$) &&&& \nl
\hfill OVRO & $54\pm8$ & $14\pm2$ & $404\pm61$ & $25\pm4$ \nl
\hfill Single Dish\tablenotemark{b} & 55 & 34 & 696 & 21 \nl
$M_{H_2,OVRO}$ ($M_\odot$)\tablenotemark{c} & $(4.9\pm0.7)\times 10^9$ &
$(1.3\pm0.2)\times 10^9$ & $(4.3\pm0.7)\times 10^9$ & $(1.8\pm0.3)\times 
10^9$ \nl
$N_{H_2, peak}$ (cm$^{-2}$) & $(6.1\pm0.9)\times10^{22}$ &
$(1.6\pm0.0.2)\times10^{22}$ & $(3.0\pm0.5)\times10^{23}$ &
$(3.7\pm0.6)\times10^{22}$ \nl
$\Sigma_{H_2}$ ($M_\odot$ pc$^{-2}$)\tablenotemark{d} &&&& \nl
\hfill peak &
$900\pm135$ & $210\pm32$ & $4670\pm700$ & $540\pm270$ \nl
\hfill mean &
$340\pm51$ & $24\pm4$ & $2650\pm400$ & $134\pm20$ \nl
$S_{1.4~GHz}$ (mJy)\tablenotemark{e} & 23.6 & 6.6 & 158 & 9.4 \nl
log $L_{1.4~GHz}$ (W Hz$^{-1}$) & 22.0 & 21.4 & 21.9 & 21.5 \nl
IRAS 12$\mu$ (Jy)\tablenotemark{f} & \multicolumn{2}{c}{($0.09\pm0.03$)} 
& $0.77\pm0.06$ & $0.12\pm0.02$ \nl
IRAS 25$\mu$ (Jy)\tablenotemark{f} & \multicolumn{2}{c}{($0.48\pm0.03$)} 
& $2.87\pm0.17$ & $0.28\pm0.02$ \nl
IRAS 60$\mu$ (Jy)\tablenotemark{f} & \multicolumn{2}{c}{($2.93\pm0.04$)} 
& $30.9\pm1.9$ & $0.83\pm0.03$ \nl
IRAS 100$\mu$ (Jy)\tablenotemark{f} & \multicolumn{2}{c}{($5.26\pm0.12$)} 
& $45.8\pm2.7$ & $1.60\pm0.11$ \nl
$T_{dust}$\tablenotemark{g} & \multicolumn{2}{c}{38 K} & 42 K & 37 K \nl
$L_{FIR}$ ($L_\odot$)\tablenotemark{g} & $2.8\times10^{10}$ &
$0.8\times10^{10}$ & $5.4\times10^{10}$ & $0.8\times10^{10}$ \nl
$L_B$ ($10^{10}L_\odot$)\tablenotemark{e} & $2.8\times10^{10}$ &
$1.6\times10^{10}$ & $3.4\times10^{10}$ & $5.5\times10^{10}$ \nl
\tablenotetext{a}{$H_\circ = 75$ km s$^{-1}$ Mpc$^{-1}$ assumed.}
\tablenotetext{b}{Single Dish references: NGC 4676: Casoli et al. (1991); 
NGC 520: Young et al. (1995); NGC 3921: Combes, personal communication.}
\tablenotetext{c}{$M_{H_2} = 1.2 \times 10^4 S_{CO}\Delta V 
D_{Mpc}^2~M_\odot$ (see \cite{San91}) }
\tablenotetext{d}{$\Sigma_{H_2}=M_{H_2} \times 
[{{\pi R^2}\over{ln2}}]^{-1}$ }
\tablenotetext{e}{From HvG96.}
\tablenotetext{f}{from IRAS ADDSCAN/SCANPI (see \cite{Helou88}).}
\tablenotetext{g}{$L_{FIR}=4\times 10^5 [2.58S_{60\mu}+S_{100\mu}] 
D_{Mpc}^2~L_\odot$ and assuming dust emissivity $n=1$ (see \cite{Helou88}).  
NGC 4676 A\&B are not resolved by IRAS, and the FIR fluxes are estimated 
by the ratio of their 1.4 GHz radio continuum flux ratio.}
\enddata
\end{deluxetable}


\bigskip

\begin{thebibliography}{}

\bibitem[Arp 1966]{Arp66} Arp, H. 1966, ApJS, 14, 1
\bibitem[Barnes 1998]{Barnes98} Barnes, J. E. 1998, in ``Galaxies:
  Interactions and Induced Star Formation'', Saas-Fee Advanced
  Course No.~26, R. C.~Kennicutt Jr., F.~Schweizer and
  J.~E.~Barnes (Springer, Berlin), p.~275.
\bibitem[Barnes \& Hernquist 1991]{BarnesH91} Barnes, J.E., \& Hernquist,
  L. 1991, ApJL, 370, L65
\bibitem[Barnes \& Hernquist 1992]{BarnesH92} Barnes, J.E., \& Hernquist,
  L. 1992, ARAA, 30, 705
\bibitem[Barnes \& Hernquist 1996]{BarnesH96} Barnes, J.E., \& Hernquist,
  L. 1996, ApJ, 471, 115
\bibitem[Binney \& Tremaine 1987]{Binney86} Binney, J. \& Tremaine, S.
  1987 in {\it ``Galactic Dynamics''}, 
  (Princeton University Press, Princeton), Chapter 2.
\bibitem[Braine et al. 2000]{Braine00} Braine, J., Lisenfeld, U., Duc, 
  P.-A., \& Leon, S. 2000, Nature, 403, 867
\bibitem[Bryant \& Scoville 1996]{Bryant96} Bryant, P.M. \& Scoville, 
  N.Z. 1996, ApJ, 457, 678
\bibitem[Bryant \& Scoville 1999]{Bryant99} Bryant, P.M. \& Scoville, 
  N.Z. 1999, AJ, 117, 2632
\bibitem[Bushouse et al. 1988]{Bush88}Bushouse, H. A., Werner, M. W.,
  \& Lamb, S. A.  1988, ApJ, 335, 74.
\bibitem[Carilli et al. 1998]{Carilli98} Carilli, C.L., Wrobel, J.M.,
  \& Ulvestad, J.S. 1998, AJ, 115, 928
\bibitem[Casoli et al. 1991]{Casoli91} Casoli, F., Boiss\'{e}, P., Combes,
F., \& Dupraz, C. 1991, A\&A, 249, 359
\bibitem[Condon et al. 1990]{Condon90} Condon, J.J., Helou, G., Sanders,
  D.B., \& Soifer, B.T. 1990, ApJS, 73, 359
\bibitem[Cram et al. 1998]{Cram98} Cram, L., Hopkins, A., Mobasher, B.,
  \& Rowan-Robinson, M. 1998, ApJ, 507, 155
\bibitem[Downes \& Solomon 1998]{Downes98} Downes, D. \& Solomon, P.M.
  1998, ApJ, 507, 615
\bibitem[Helou et al. 1988]{Helou88} Helou, G., Khan, I.R., Malek, L.,
  \& Boehmer, L. 1988, ApJS, 68, 151
\bibitem[Hibbard et al. 1994]{Hib94} Hibbard, J.E., Guhathakurta, P.,
  van Gorkom, J.H., \& Schweizer, F. 1994, AJ, 107, 67
\bibitem[Hibbard \& Mihos 1995]{Hib95b} Hibbard, J.E., \&
  Mihos, J.C. 1995, AJ, 110, 140
\bibitem[Hibbard \& van Gorkom 1996]{Hib96} Hibbard, J.E., \&
  van Gorkom, J.H. 1996, AJ, 111, 655 (HvG96)
\bibitem[Kenney et al. 1992]{Kenney92} Kenney, J.D.P., Wilson, C.D., 
  Scoville, N.Z., Devereux, N.A., \& Young, J.S. 1992, ApJ, 395, 79
\bibitem[Kennicutt 1989]{Kennicutt89} Kennicutt, R. C.~Jr. 1989, ApJ, 
  344, 685
\bibitem[Kennicutt 1998]{Kennicutt98b} Kennicutt, R. C.~Jr. 1998, 
  ARAA, 36, 189
\bibitem[Lake \& Dressler 1986]{Lake86} Lake, G., \& Dressler, A. 1986,
  ApJ, 310, 605
\bibitem[Landolt 1983]{Landolt83} Landolt, A. U. 1983, AJ, 88, 439
\bibitem[Landolt 1992]{Landolt92} Landolt, A. U. 1992, AJ, 104, 340
\bibitem[Lo et al. 1987]{Lo87} Lo, K.Y., Cheung, K.W., Masson, C.R.,
  Phillips, T.G., Scott, S.L., Woody, D.P. 1987, ApJ, 312, 574
\bibitem[Mihos et al. 1993]{Hos93} Mihos, J.C., Bothun, G.D., \&
  Richstone, D.O. 1993, ApJ, 418, 82
\bibitem[Mihos \& Hernquist 1996]{Hos96} Mihos, J.C., \& Hernquist,
  L. 1996, ApJ, 464, 641
\bibitem[Negroponte \& White 1983]{Negroponte83} Negroponte, J., White, 
  S.D.M. 1983, MNRAS, 205, 1009
\bibitem[Olson \& Kwan 1990]{Olson90} Olson, K.M., Kwan, J. 1990, 
  ApJ, 361, 426
\bibitem[Sanders et al. 1984]{San84} Sanders, D.B., Solomon, P.M.,
  \& Scoville, N.Z. 1984, ApJ, 276, 182
\bibitem[Sanders et al. 1988]{San88b} Sanders, D.B., Scoville, N.Z.,
  Sargent, A.I., Soifer, B.T. 1988, ApJL, 324, L55
\bibitem[Sanders et al. 1991]{San91} Sanders, D.B., Scoville, N.Z.,
  \& Soifer, B.T. 1991, ApJ, 370, 158
\bibitem[Sanders \& Mirabel 1996]{San96} Sanders, D.B., 
  \& Mirabel, I.F. 1996, ARAA, 34, 749
\bibitem[Schweizer 1982]{Schwe82} Schweizer, F. 1982, ApJ, 252, 455
\bibitem[Schweizer 1996]{Schwe96a} Schweizer, F. 1996, AJ, 111, 109 (S96)
\bibitem[Schweizer et al. 1996]{Schwe96b} Schweizer, F., Miller, B. W.,
  Whitmore, B.C., \& Fall, S.M. 1996, AJ, 112, 1839
\bibitem[Scoville et al. 1992]{Sco92} Scoville, N.Z., Carlstrom,
  J.C., Chandler, C.J., Phillips, J.A., Scott, S.L., Tilanus,
  R.P., \& Wang, Z. 1992, PASP, 105, 1482
\bibitem[Scoville et al. 1997]{Sco97} Scoville, N.Z., Yun, M.S., \&
  Bryant, P.M. 1997, ApJ, 484, 702
\bibitem[Shen \& Lo 1995]{Shen95} Shen, J.J. \&
  Lo, K.Y. 1995, ApJ, 450, L39
\bibitem[Shepherd et al. 1994]{Shepherd94} Shepherd, M.C., Pearson, 
  T.J., \& Taylor, G.B. 1994, BAAS, 26, 987
\bibitem[Smith \& Higdon 1994]{Smith94} Smith, B. J., \& Higdon, J. L. 
  1994, AJ, 108, 837
\bibitem[Smith et al. 1999]{Smith99} Smith, B. J., Struck, C., Kenney,
  J. D. P. \& Jogee, S. 1999, AJ, 117, 1237
\bibitem[Solomon et al. 1992]{Sol92} Solomon, P. M., Downes, D., \&
  Radford, S. J. E. 1992, ApJ, 387, L55
\bibitem[Stanford \& Balcells 1990]{StanfordB90} Stanford, S.A., \&
  Balcells, M. 1990, ApJ, 355, 59
\bibitem[Stanford et al. 1990]{Stanford90} Stanford, S.A., Sargent, 
  A.I., Sanders, D.B., \& Scoville, N.Z. 1990, ApJ, 349, 492
\bibitem[Stockton 1974]{Stockton74} Stockton, A. 1974, ApJ, 187, 219
\bibitem[Stockton \& Bertola 1980]{Stockton80} Stockton, A., \& Bertola,
  F. 1980, ApJ, 235, 37
\bibitem[Thronson et al. 1989]{Thronson89} Thronson, H. A., Tacconi, L.,
	Kenney, J., Greenhouse, M. A., Margulis, M. et al. 1989, ApJ, 344, 747
\bibitem[Toomre 1977]{Toomre77} Toomre, A. 1977, in {\it The Evolution of
  Galaxies and Stellar Populations,} eds. B.M. Tinsley \& R.B.
  Larson (New Haven: Yale University), p.401
\bibitem[Toomre \& Toomre 1972]{Toomre72} Toomre, A., \& Toomre, J. 
  1972, ApJ, 178, 623
\bibitem[Wang et al. 1992]{Wang92} Wang, Z., Schweizer, F., \& Scoville, 
  N.Z. 1992, ApJ, 396, 510
\bibitem[Wainscoat \& Cowie 1992]{Wainscoat92} Wainscoat, R. J. \& Cowie, 
  L. L. 1992, AJ, 103, 332
\bibitem[Wilson et al. 2000]{Wilson00} Wilson, C. D., Scoville, N.,
  Madden, S. C. \& Charmandaris, V. 2000, ApJ, 542, 120
\bibitem[Young \& Knezek 1989]{Young89} Young, J. S., \& Knezek, P. M. 1989,
	ApJ, 347, L55
\bibitem[Young \& Scoville 1991]{Young91} Young, J.S., \& Scoville, N.Z. 
  1991, ARAA, 29, 581
\bibitem[Young et al. 1995]{Young95} Young, J.S., Xie, S., Tacconi, L.,
  Knezek, P., Viscuso, P. et al. 1995, ApJS, 98, 219
\bibitem[Yun \& Hibbard 2000]{PaperII} Yun, M.S., \& Hibbard, J. E. 2000, 
  ApJ, submitted (Paper~II)

\end{thebibliography}
\end{document}